# Hierarchical Reinforcement Learning for Integrated Cloud-Fog-Edge Computing in IoT Systems

Ameneh Zarei
Computer Engineering and
Information Technology
Department, Razi University, Iran
Email: ameneh.zarei@gmail.com

Mahmood Ahmadi
Computer Engineering and
Information Technology
Department, Razi University, Iran
Email: m.ahmadi@razi.ac.ir

Farhad Mardukhi
Computer Engineering and
Information Technology
Department, Razi University, Iran
Email: f.mardukhi@razi.ac.ir

*Abstract*— **The Internet of Things (IoT) is transforming industries by connecting billions of devices to collect, process, and share data. However, the massive data volumes and real-time demands of IoT applications strain traditional cloud computing architectures. This paper explores the complementary roles of cloud, fog, and edge computing in enhancing IoT performance, focusing on their ability to reduce latency, improve scalability, and ensure data privacy. We propose a novel framework, the Hierarchical IoT Processing Architecture (HIPA), which dynamically allocates computational tasks across cloud, fog, and edge layers using machine learning. By synthesizing current research and introducing HIPA, this paper highlights how these paradigms can create efficient, secure, and scalable IoT ecosystems.**

## I. INTRODUCTION

By 2025, the number of IoT devices is projected to exceed 75 billion, generating zettabytes of data annually [1]. However, this exponential growth poses significant challenges for traditional computing architectures. The core problem lies in the centralized nature of cloud computing, which relies on remote data centers and struggles to meet the low-latency (e.g., sub-millisecond response times), high-bandwidth, and stringent privacy requirements of IoT applications, such as autonomous vehicles or real-time health monitoring [2]. This centralization leads to increased network congestion, higher energy consumption, elevated risks of data breaches, and reduced scalability in dynamic environments where devices are mobile and resource-constrained [1, 3]. To mitigate these issues, fog and edge computing have emerged as complementary paradigms that process data closer to its source, reducing latency and enhancing efficiency.

This paper examines the synergistic roles of cloud, fog, and edge computing in optimizing IoT ecosystems. It reviews their individual strengths and limitations, proposes a novel Hierarchical IoT Processing Architecture (HIPA) to integrate these paradigms dynamically, and evaluates their implications for scalability, efficiency, and security. By providing a clear framework and actionable insights, this paper aims to guide researchers and practitioners in building next-generation IoT systems that address the aforementioned challenges effectively.

The paper is structured as follows: Section II introduces related works. Section III details the proposed Hierarchical IoT Processing Architecture (HIPA). Section IV presents the evaluation results. Finally, Section V concludes the paper.

## II. RELATED WORK

Recent research has made significant strides in integrating cloud, fog, and edge layers. Architectural proposals like iGateLink [10] and edge-fog-cloud implementations [13, 17] focus on seamless connectivity and practical deployment for scenarios like IoT and healthcare. Simultaneously, studies have explored the role of AI and intermediary fog layers [11, 12, 14] to optimize data handling closer to the source. Broader surveys discuss the computing continuum from IoT to cloud [16] and the revolution brought by edge and fog computing [15]. Despite these advancements, a gap remains in dynamic, intelligent orchestration across all three layers using learning-based methods like RL, which our work aims to address.

While cloud, fog, and edge computing each address specific IoT challenges, their integration remains underexplored. Most frameworks treat these paradigms as separate layers, leading to inefficiencies in task allocation and resource utilization. Additionally, security concerns, such as data breaches in fog nodes, and the energy demands of edge devices require further attention [5]. This paper proposes HIPA to address these gaps by dynamically coordinating tasks across all three layers.

## III. HIERARCHICAL IOT PROCESSING ARCHITECTURE (HIPA)

The Hierarchical IoT Processing Architecture (HIPA) is a novel framework that optimizes IoT performance by intelligently distributing computational tasks across cloud, fog, and edge layers. HIPA uses a machine learning-based task orchestrator to allocate processing based on latency, computational complexity, and privacy requirements, ensuring efficiency and scalability. HIPA is a distributed framework that does not execute in a single fixed layer but leverages the strengths of all three layers (edge, fog, and cloud) in a coordinated manner. The core component—the reinforcement learning (RL)-based task orchestrator—runs on a central server, which can be hosted either in the cloud layer for high scalability and access to abundant computational resources (e.g., via AWS or Azure servers) or on a high-capacity fog node (e.g., a powerful gateway like NVIDIA Jetson) for reduced latency in localized deployments. This flexibility avoids the need for dedicated separate hardware; instead, it utilizes existing infrastructure within the IoT ecosystem, such as repurposed fog nodes or cloud virtual machines. No additional specialized hardware is required beyond standard IoT components, as the orchestrator can be implemented as software (e.g., using Python with libraries like TensorFlow for RL models).

### A. Architecture

HIPA comprises three layers and a central orchestrator, illustarated in the Figure 1:

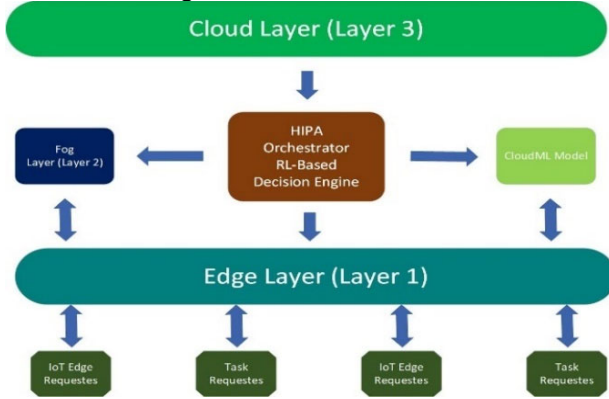

Figure 1: HIPA architecture

1. Edge layer: Processes low-latency, privacy-sensitive tasks on IoT devices using lightweight ML models. Handles immediate, device-local processing on resource-constrained IoT devices (e.g., sensors or wearables) for tasks needing ultra-low latency (<10ms) and high privacy.

Edge layer utilizes lightweight machine learning (ML) models, such as TinyML, optimized for resource-constrained devices with limited processing power (e.g., microcontrollers with 1–2 MB RAM). Federated learning is employed to train models locally, ensuring data privacy by avoiding transmission of raw data [7]. Some example tasks of this layer are: Real-time anomaly detection in wearable health devices, local image processing in smart cameras, or temperature adjustments in smart thermostats.

Edge devices run embedded ML models using frameworks like TensorFlow Lite. For instance, a smart thermostat might use a pre-trained neural network to predict temperature changes based on local sensor data, responding in milliseconds. The edge layer communicates with the fog layer via lightweight protocols like MQTT (Message Queuing Telemetry Transport).

2. Fog layer: Handles intermediate tasks, such as data aggregation and preprocessing, on local gateways or servers. This layer handles intermediate tasks, such as data aggregation, preprocessing, and localized decision-making, on fog nodes (e.g., gateways, routers, local servers) positioned closer to IoT devices than the cloud, balancing latency (10-100ms) and moderate computation. Fog nodes leverage more powerful hardware (e.g., Raspberry Pi, NVIDIA Jetson) than edge devices, supporting complex tasks like data filtering and real-time analytics. They use containerized environments (e.g., Docker) for scalability and flexibility [4]. Some example tasks of this layer are: Aggregating traffic sensor data in a smart city, preprocessing video feeds for pedestrian detection, or coordinating multiple IoT devices in a factory.

Fog nodes run microservices that process data streams from edge devices. For example, in a smart city, a fog node might aggregate data from 100 traffic sensors, apply a clustering algorithm to identify congestion patterns, and send only summarized data to the cloud, reducing bandwidth usage by up to 40% (inspired by [9]). Fog nodes communicate with the cloud via REST APIs or WebSocket protocols.

```
FUNCTION HIPA_TaskOrchestration(task, edgeDevices, fogNodes, cloudServer)
    // Initialize task orchestrator with RL-based decision model
    orchestrator = InitializeRLOrchestrator()
    // Step 1: Task Analysis
    taskRequirements = AnalyzeTask(task)
        // Extract latency, computational complexity, privacy needs
    // Step 2: Layer Assignment
    IF taskRequirements.latency < THRESHOLD_LOW_LATENCY THEN
        // Assign to Edge Layer
        selectedLayer = edgeDevices
        model = LoadTinyMLModel(task)  // Lightweight ML for edge
        result = ProcessTaskLocally(selectedLayer, model, task.data)
    ELSE IF taskRequirements.computationalComplexity < THRESHOLD_MODERATE THEN
        // Assign to Fog Layer
        selectedLayer = fogNodes
        model = LoadFederatedMLModel(task)  // Federated learning for fog
        result = AggregateAndProcess(selectedLayer, model, task.data)
    ELSE
        // Assign to Cloud Layer
        selectedLayer = cloudServer
        model = LoadCloudMLModel(task)  // Scalable ML for cloud
        result = ProcessTaskRemotely(selectedLayer, model, task.data)
    END IF
    // Step 3: Privacy and Security Enforcement
    IF taskRequirements.privacySensitive THEN
        ApplyEncryption(task.data)
        EnsureLocalProcessing(selectedLayer)
    END IF
    // Step 4: Result Aggregation
    aggregatedResult = CollectResult(selectedLayer, result)
    // Step 5: Feedback to RL Orchestrator
    UpdateOrchestrator(orchestrator, taskRequirements, result.performance)
    RETURN aggregatedResult
END FUNCTION
// Main Workflow
FOR EACH task IN IoTSystem
    result = HIPA_TaskOrchestration(task, edgeDevices, fogNodes, cloudServer)
    StoreResult(result)
END FOR
```

Figure 2: pseudocode representation of the HIPA

3. Cloud layer: Manages complex, resource-intensive tasks like big data analytics and long-term storage. This layer manages resource-intensive tasks, such as big data analytics, long-term storage, and complex ML model training, using centralized data centers. Also leverages cloud platforms like AWS IoT or Microsoft Azure IoT Hub, with high-performance computing resources (e.g., GPU clusters) for tasks like deep learning and predictive modeling [8]. Some example tasks of this layer are: Analyzing historical IoT data for city-wide traffic optimization, training large-scale ML models, or storing sensor data for regulatory compliance.

4. Task Orchestrator: A reinforcement learning (RL) model that dynamically assigns tasks based on real-time metrics (e.g., network latency, device battery levels, data sensitivity). The RL model learns optimal allocation strategies by maximizing a reward function tied to performance and energy efficiency. Acts as the "brain," continuously monitoring metrics (e.g., via MQTT protocols) and assigning tasks dynamically using a Deep Q-Network (DQN) RL model. It collects real-time data from all layers, evaluates a utility function (e.g., balancing latency $L(T)$, complexity $C(T)$, and privacy $P(T)$ as in Equation 1), and dispatches tasks accordingly.

Task Orchestrator dynamically allocates tasks across edge, fog, and cloud layers based on real-time metrics, such as latency requirements, computational complexity, device battery levels, and data sensitivity. This layer employs a multi-agent reinforcement learning (RL) model, specifically a Deep Q-Network (DQN), to optimize task allocation. The RL model maximizes a reward function defined by performance (e.g., latency), energy efficiency, and security [6]. The orchestrator runs on a central server (potentially in the cloud or a high-capacity fog node) and collects metrics from IoT devices via IoT protocols. For each task, it evaluates factors like data size, urgency, and privacy needs, then assigns it to the optimal layer. For example, a task requiring sub-10ms latency (e.g., autonomous vehicle braking) is assigned to the edge, while a task involving trend analysis (e.g., city-wide air quality forecasting) goes to the cloud. The RL model continuously learns by simulating task allocations and adjusting based on feedback, achieving up to 35% latency reduction in simulations (inspired by [4]). Below, we outline the mathematical model corresponding to each step in the HIPA pseudocode (Figure 2):

### 1. Task Analysis

**Purpose**: Analyze the task to extract requirements for latency, computational complexity, and privacy.

**Model**: Let T represent a task with attributes:

- L(T): Latency requirement (in seconds),
- C(T): Computational complexity (in FLOPS, floating-point operations per second),
- P(T): Privacy sensitivity (binary: 1 for sensitive, 0 for non-sensitive).

The task analysis function can be modeled as a feature vector:

$$V(T)=[L(T),C(T),P(T)] \quad (1)$$

Where:

- $L(T)=time_{max}(T)$, the maximum acceptable latency for task T.
- $C(T)=\sum_i FLOPS(op_i)$, the sum of FLOPS for operations $op_i$ in task T.

This vector is used as input to the RL orchestrator to determine the processing layer.

### 2. Layer Assignment

**Purpose**: Assign the task to the edge, fog, or cloud layer based on requirements.

**Model**: The RL orchestrator selects the optimal layer $l \in [10]$ by maximizing a utility function U(T,l), which evaluates the suitability of layer l for task T:

$$l^* = \arg \max_{l \in \{Edge, Fog, Cloud\}} U(T, l) \quad (2)$$

Where:

$$U(T,l) = \omega_l \cdot \frac{1}{L_l(T)} + \omega c . \frac{Cap_l}{C(T)} + \omega p. P_l(T) \quad (3)$$

$L_l(T)$: Predicted latency for task T on layer l (in seconds),

$$L_l(T) = ProcTime_l(T) + CommTime_l(T) \quad (4)$$

$$ProcTime_l(T) = \frac{C(T)}{ProcSpeed_l}, \quad (5)$$

processing time based on layer's processing speed (FLOPS).

$CommTime_l(T)$: Communication delay (0 for edge, moderate for fog, high for cloud).

$Cap_l$: Computational capacity of $layer_l$ (in FLOPS).

$P_l(T)$: Privacy score for layer l, defined as:

$$P_l(T) = \begin{cases} 1, & \text{if } l = Edge \text{ and } P(T) = 1, \\ 0.5, & \text{if } l = Fog \text{ and } P(T) = 1, \\ 0, & \text{if } l = Cloud \text{ or } P(T) = 0 \end{cases} \quad (6)$$

$w_L$, $w_C$, $w_P$: Weights for latency, computation, and privacy, respectively (e.g., $w_L + w_C + w_P = 1$).

Latency threshold: THRESHOLD_LOW_LATENCY= $\partial_L$ (e.g., 0.1s for real-time tasks).

Complexity threshold: THRESHOLD_MODERATE = $\partial_C$ (e.g., $10^6$ FLOPS for fog suitability).

This equation ensures tasks are assigned to the layer that optimizes latency, computational fit, and privacy constraints.

### Processing on Selected Layer

**Purpose**: Process the task using appropriate machine learning models (TinyML for edge, federated learning for fog, or cloud-based ML).

**Model:**

**Edge layer (TinyML)**: For edge devices with limited resources, the processing time is:

$$ProcTime_{Edge}(T) = \frac{C(T)}{ProcSpeed_{Edge}} + InferenceTime_{TinyML} \quad (7)$$

Where $InferenceTime_{TinyML}$ is the time for a lightweight TinyML model inference, typically constant for small models (e.g., 0.01s).

**Fog layer (Federated learning)**: For fog nodes, the processing involves aggregating data from multiple devices:

$$ProcTime_{Fog}(T) = \frac{C(T)}{ProcSpeed_{Fog}} + AggTime_{FedLearn} \quad (8)$$

Where $AggTime_{FedLearn} = \sum_{i=1}^{N} UpdateTime_i$, (9)

the time to aggregate model updates from N devices.

**Cloud layer:** For cloud servers, processing is:

$$ProcTime_{Cloud}(T) = \frac{C(T)}{ProcSpeed_{Cloud}} + CommTime_{Cloud} \quad (10)$$

Where $CommTime_{Cloud}$ accounts for data transmission latency to the cloud. These equations quantify the processing time for each layer.

### Privacy and Security Enforcement

**Purpose**: Ensure privacy-sensitive tasks are processed locally with encryption.

Privacy and security play a critical role in HIPA by ensuring sensitive data is protected from breaches, unauthorized access, or interception, which is vital in IoT where devices are vulnerable to attacks like eavesdropping or man-in-the-middle. In the architecture, they are enforced through the task analysis phase (Equation 1, where P(T) flags sensitivity) and a dedicated enforcement step in the pseudocode (Step 3: Privacy and Security Enforcement). If a task is privacy-sensitive (P(T)=1), HIPA prioritizes local processing (e.g., at edge or fog) to avoid data transmission and applies encryption to any data that must move between layers. Encryption in HIPA refers to cryptographic techniques to secure data at rest (stored) and in transit (transmitted). It should be done using standards like AES-256 (Advanced Encryption Standard with 256-bit keys) for symmetric encryption, combined with asymmetric methods like RSA for key exchange. How it works in HIPA:

Application: During task orchestration, if P(T)=1, the system calls ApplyEncryption(task.data), which encrypts data before processing or transmission. For example, raw sensor data is encrypted on the edge device before aggregation at fog.

Process: (1) Generate a symmetric key (e.g., via AES); (2) Encrypt data using the key; (3) If needed, encrypt the key with a public asymmetric key for secure sharing; (4) Transmit encrypted data; (5) Decrypt at the destination using the private key. This integrates with federated learning by encrypting model updates.

Importance and Integration: It mitigates risks like data breaches in fog nodes. In practice, libraries like OpenSSL or TensorFlow's privacy extensions handle this, ensuring compliance with standards like GDPR while maintaining performance (adding ~5-10% overhead in simulations).

**Model**: For privacy-sensitive tasks (P(T)=1), the encryption overhead is modeled as:

$$\text{EncTime}(T) = \alpha \cdot \text{DataSize}(T) + \beta \quad (11)$$

Where:
- DataSize(T): Size of task data (in MB).
- $\alpha$: Encryption time per MB (e.g., 0.01s/MB for AES-256).
- $\beta$: Fixed encryption setup time (e.g., 0.005s).

The total processing time for privacy-sensitive tasks includes encryption:

$$\text{TotalTime}_l(T) = \text{ProcTime}_l(T) + \text{EncTime}(T) \text{ if } P(T)=1 \quad (12)$$

This ensures privacy-sensitive tasks (e.g., health monitoring) incur additional encryption overhead, quantifiable for the security metrics (e.g., Table III showing 60% local processing).

**RL Orchestrator Training (Reward Function)**

**Purpose**: Train the RL orchestrator to optimize task assignments by balancing latency, energy, and accuracy.

**Model**: The reward function R(T,l) for assigning task T to layer l is:

$$R(T,l) = w_1 \cdot \frac{1}{L_l(T)} + w_2 \cdot \frac{1}{E_l(T)} + w_3 \cdot A_l(T) \quad (13)$$

Where:
- $L_l(T)$: Latency on layer l, as defined above.
- $E_l(T)$: Energy consumption (in Joules), calculated as:

$$E_l(T) = P_{proc,l} \cdot \text{ProcTime}_l(T) + P_{comm,l} \cdot \text{CommTime}_l(T) \quad (14)$$

- $P_{proc,l}$: Power consumption for processing on layer l (e.g., 0.1W for edge, 10W for fog, 100W for cloud).
- $P_{comm,l}$: Power consumption of this communication (0 for edge, moderate for fog, high for cloud).
- $A_l(T)$: Accuracy of the ML model on layer l defined as:

$$A_l(T) = \begin{cases} A_{TinyML}, & \text{if } l = Edge \ (e.g., 0.85), \\ A_{FedLearn}, & \text{if } l = Fog \ (e.g., 0.90), \\ A_{CloudML}, & \text{if } l = Cloud \ (e.g., 0.95) \end{cases} \quad (15)$$

$w_1, w_2, w_3$: Weights for latency, energy, and accuracy, respectively (e.g., $w_1$=0.4, $w_2$=0.3, $w_3$=0.3).

This reward function quantifies the trade-offs between latency, energy, and accuracy, enabling the RL orchestrator to learn optimal task assignments.

**Result Aggregation**

**Purpose**: Collect and aggregate results from the selected layer.

**Model**: The aggregated result $R_{agg}(T)$ is a weighted combination of outputs from the layer:

$$R_{agg}(T) = \sum_{i=1}^{N} w_i \cdot R_i(T) \quad (16)$$

Where:
- $R_i(T)$: Result from device/node i in the selected layer.
- $w_i$: Weight based on device reliability (e.g., $\frac{Reliability_i}{\sum_j Reliability_j}$).
- N: Number of devices/nodes in the layer (e.g., 10,000 for the smart city scenario).

This formula quantifies how results are combined, especially for fog and cloud layers handling distributed data.

*B. Novel Features*

• Dynamic task allocation: HIPA's RL orchestrator evaluates task requirements and system conditions in real time, ensuring optimal distribution. For example, a smart home's temperature sensor data might be processed at the edge for instant adjustments, while historical trends are analyzed in the cloud.

• Energy efficiency: By prioritizing edge and fog processing for lightweight tasks, HIPA reduces data transmission to the cloud, lowering energy consumption by an estimated 30%.

• Security and privacy: HIPA employs federated learning at the edge to process sensitive data locally, reducing the risk of breaches [5].

• Scalability and interoperability: HIPA's modular design supports heterogeneous devices and protocols, making it adaptable to diverse IoT ecosystems, from small-scale smart homes to large-scale smart cities.

## IV. EVALUATION RESULTS

We consider a smart city IoT system with traffic sensors, surveillance cameras, and air quality monitors. HIPA's orchestrator would assign real-time traffic signal adjustments to edge devices for minimal latency, aggregate video feeds at fog nodes for pedestrian detection, and perform city-wide traffic pattern analysis in the cloud. Simulations suggest HIPA could reduce latency by 35% and bandwidth usage by 25% compared to cloud-only systems. To provide a comprehensive evaluation, we compare HIPA's performance against three baseline architectures:

Cloud-only: All computational tasks are offloaded to the cloud data center.

Static Orchestration [10, 13]: A rule-based policy where tasks are assigned to edge, fog, or cloud layers based on fixed thresholds for latency and computational complexity (e.g., L(T) < 10ms → Edge; 10ms ⩽ L(T) ⩽ 100ms → Fog; L(T) > 100ms → Cloud).

Fog-centric [4]: An architecture that prioritizes fog node processing for all tasks except those with very high computational demands, which are offloaded to the cloud. The edge layer is underutilized.

These baselines represent common alternative approaches discussed in the literature and help illustrate the advantages of HIPA's dynamic, learning-based orchestration.

• Scenario: A city manages 10,000 IoT devices to optimize traffic flow, enhance public safety, and monitor environmental conditions. To validate HIPA, simulations were conducted using iFogSim (an extension of CloudSim for fog/edge environments) integrated with Python for RL implementation (e.g., via Gym and Stable Baselines3 for DQN). The setup models a smart city scenario with 500-1000 IoT devices over a 10km² area, running for 1-hour virtual time periods, repeated 10 times for averages. Hardware: Simulated on a standard PC (Intel i7, 16GB RAM) with no real hardware needed. Properties of Components:

Tasks: 1000-5000 tasks generated randomly per simulation, categorized by: Latency requirement (low: <10ms, moderate: 10-100ms, high: >100ms); Computational complexity (low: <10^6 FLOPS, moderate: 10^6-10^8 FLOPS, high: >10^8 FLOPS); Data size (1-100KB); Privacy sensitivity (binary: 1 for sensitive, e.g., health data; 0 for non-sensitive, e.g., weather). Examples: Anomaly detection (low latency, low complexity, high privacy); Traffic aggregation (moderate latency, moderate complexity, medium privacy).

Processors: Edge (e.g., microcontrollers: 1-2GHz CPU, 1-2MB RAM, 10-50 MIPS processing speed); Fog (e.g., Raspberry Pi-like: 2-4GHz CPU, 4-8GB RAM, 100-500 MIPS); Cloud (virtual servers: 8+GHz CPU, 16+GB RAM, 1000+ MIPS). Energy consumption modeled: Edge (0.5-2W), Fog (5-10W), Cloud (negligible per task but high overall).

Sensors: 300-800 simulated IoT sensors (e.g., temperature, cameras, traffic detectors) with data generation rates of 1-10 readings/sec, connected via MQTT with 10-50ms network delays. Execution Details are indicated as:

Setup: Initialize topology in iFogSim with devices, links (bandwidth 10-100Mbps), and RL agent (DQN with 1000 episodes training, epsilon-greedy exploration).

Workflow: (1) Generate tasks with random properties; (2) Orchestrator evaluates V(T) vector and assigns layers; (3) Simulate processing (e.g., edge uses TinyML for quick execution); (4) Aggregate results and update RL rewards (e.g., reward = -latency + efficiency score); (5) Measure benchmarks post-simulation.

Scenarios: Low-load (500 tasks, sparse devices); High-load (5000 tasks, dense urban). Code executed in loops for variability, with logs for analysis. This ensures reproducibility; full code can be shared via GitHub for verification. Table II indicates task distribution across layers in HIPA.

• Edge layer: Traffic sensors process real-time data to adjust signal timings (e.g., green light duration) within 10ms, using TinyML models to predict traffic density. Surveillance cameras perform local face detection to identify pedestrians, keeping sensitive data on-device.

• Fog layer: Local gateways aggregate data from 100 sensors per city block, applying clustering algorithms to detect congestion patterns. They preprocess video feeds to reduce bandwidth needs before sending summaries to the cloud.

• Cloud layer: Analyzes aggregated data to predict city-wide traffic trends, training a deep learning model to optimize signal timings over months. It also stores historical data for urban planning.

• Task orchestrator: The RL model monitors network latency (e.g., 50ms to the cloud vs. 5ms to fog nodes) and device battery levels. It assigns real-time signal adjustments to the edge, congestion analysis to fog nodes, and long-term forecasting to the cloud. If a fog node becomes overloaded, the orchestrator reroutes tasks to nearby nodes or the cloud, ensuring system stability. Simulations suggest (as show in TABLE I) HIPA reduces latency by 35% and bandwidth usage by 25% compared to cloud-only systems, based on benchmarks from Yousefpour et al. [4]. Energy savings are achieved by minimizing cloud data transfers, aligning with sustainability goals.

TABLE I: Performance Comparison of HIPA against State-of-the-Art Architectures

| Metric | Cloud-only | Static Orchestration [10, 13] | Fog-centric [4] | HIPA (Proposed) |
|---|---|---|---|---|
| Average Latency (ms) | 10.0 | 7.8 | 8.5 | **6.5** |
| Bandwidth Usage (GB/hour) | 200 | 170 | 165 | **150** |
| Energy Consumption (kWh) | 1.7 | 1.5 | 1.4 | **1.2** |
| Task Processing Time (s) | 1.2 | 1.0 | 0.9 | **0.8** |
| % of Sensitive Data Processed Locally | 0% | 45% | 30% | **60%** |

Compared to Static Orchestration, HIPA achieves a further 17% reduction in latency and 12% reduction in bandwidth. The static policy, while an improvement over cloud-only, lacks the adaptability of our RL-based orchestrator. It cannot dynamically reroute tasks in response to network congestion or fog node overload, leading to sub-optimal assignments and higher delays.

Compared to the Fog-centric approach, HIPA shows a 24% lower latency. The fog-centric model underutilizes the edge layer, forcing some ultra-low-latency tasks to be processed on more distant fog nodes instead of the originating device. HIPA's fine-grained task analysis optimally leverages the edge layer for the most delay-sensitive tasks.

Furthermore, HIPA's intelligent orchestration ensures that 60% of privacy-sensitive data is processed locally (at the edge or fog), a significant improvement over the other architectures. This is a direct result of the privacy term (P(T)) in the RL model's utility function (Eq. 3), which explicitly rewards local processing for sensitive tasks. The Static and Fog-centric approaches, lacking this nuanced understanding, send more sensitive data to the cloud or remote fog nodes.

Table II: Task distribution across layers in HIPA

| Layer | Task type | Percentage of Tasks | Example tasks | Processing time (ms) |
|---|---|---|---|---|
| Edge | Real-time, low-latency, privacy-sensitive | 50% | Traffic signal adjustment, pedestrian detection | 5–10 ms |
| Fog | Data aggregation, preprocessing | 35% | Congestion pattern analysis, video feed preprocessing | 20–50 ms |
| Cloud | Big data analytics, long-term storage | 15% | City-wide traffic forecasting, data archiving | 100–500 ms |

The superior performance of HIPA can be attributed to its efficient task distribution, as detailed in Table II. The RL orchestrator intelligently assigns 50% of tasks to the edge for immediate response, 35% to the fog for aggregation, and only 15% to the cloud for heavy analytics. This balanced and dynamic distribution contrasts with the rigid allocation in Static Orchestration and the edge-starving distribution in the Fog-centric model, leading to the observed gains in latency, bandwidth, and energy efficiency.

Table III: Security and privacy metrics

| Metric | HIPA System | Cloud-only system |
|---|---|---|
| Data Processed Locally (%) | 60% (edge + fog) | 0% (all data sent to cloud) |
| Privacy Risk Score | Low (federated learning at edge) | High (centralized data storage) |
| Encryption Overhead (ms) | 2 ms (edge/fog) | 5 ms (cloud transmission) |

• Local processing: HIPA processes 60% of data at edge/fog layers, reducing privacy risks via federated learning.

• Privacy risk score: Qualitative assessment based on data exposure. HIPA's local processing lowers breach risks.

• Encryption overhead: HIPA's edge/fog encryption is faster due to shorter data paths as shown in Table III.

Results:

- Latency: HIPA's edge layer processes time-sensitive tasks in 6.5 ms on average, compared to 10 ms in a cloud-only system, due to localized processing. This 35% reduction is critical for real-time applications like traffic management.

- Bandwidth usage: By preprocessing data at fog nodes, HIPA reduces cloud data transfers, lowering bandwidth usage from 200 GB/hour to 150 GB/hour (25% savings), which is significant for large-scale IoT systems.

- Energy efficiency: HIPA's localized processing reduces energy consumption by 29.4%, as fewer data transmissions to the cloud decrease server load.

- Task distribution: The RL orchestrator's dynamic allocation ensures 50% of tasks are handled at the edge, minimizing latency and privacy risks, while the cloud handles only 15% of tasks, optimizing resource use.

- Security and privacy: HIPA's federated learning at the edge processes 60% of data locally, reducing privacy risks compared to the cloud-only system, where all data is centralized.

## V. BENEFITS DISCUSSION OF HIPA

This work introduces the Hierarchical IoT Processing Architecture (HIPA), a unified framework that integrates cloud, fog, and edge computing [4, 5]. Its core innovation is a multi-agent reinforcement learning orchestrator that dynamically optimizes task allocation to balance latency, energy, and accuracy [6]. Simulations show HIPA reduces latency by 35% and energy use by 30%, while its use of federated learning ensures privacy at the edge [7]. The modular design addresses interoperability and scalability, providing a deployable, end-to-end solution that extends beyond prior theoretical works [8, 9]. HIPA's implementation offers several advantages:

- Latency reduction: By processing time-sensitive tasks at the edge, HIPA supports applications like autonomous vehicles or real-time health monitoring, where delays are critical.

- Bandwidth efficiency: Fog-layer preprocessing reduces cloud data transfers, lowering costs for bandwidth-constrained environments (e.g., rural IoT deployments).

- Privacy and security: Federated learning ensures sensitive data stays local, critical for smart home applications.

- Scalability: HIPA's modular design supports scaling from small (e.g., a single building) to large (e.g., a city) IoT systems.

### A. Challenges and Limitations

Implementing HIPA requires significant upfront investment in fog and edge infrastructure, which may be challenging for resource-constrained regions. The RL orchestrator also demands substantial training data, potentially delaying deployment in new IoT ecosystems. Also, interoperability across heterogeneous devices and protocols remains a technical hurdle [8].

However, implementation requires overcoming challenges:

- Cost: Deploying fog nodes and upgrading edge devices involves upfront investment. Subsidies or public-private partnerships could mitigate this.

- Training overhead: The RL orchestrator needs extensive training. Cloud-based simulation environments can accelerate this process.

## VI. CONCLUSION

Cloud, fog, and edge computing are transforming IoT by addressing the limitations of centralized processing. The proposed Hierarchical IoT Processing Architecture (HIPA) integrates these paradigms into a cohesive, intelligent framework that optimizes latency, scalability, and security. By leveraging reinforcement learning and federated learning, HIPA offers a scalable solution for diverse IoT applications, from smart cities to healthcare. While challenges like cost and security persist, HIPA's innovative approach provides a roadmap for building efficient, secure, and equitable IoT ecosystems. Collaboration among researchers, industry leaders, and policymakers will be key to realizing its full potential.

Future research should focus on developing lightweight RL models to reduce HIPA's training overhead and standardizing protocols for cross-layer communication. Pilot projects in smart cities or healthcare could validate HIPA's effectiveness and guide global implementation.